\documentclass{aastex}
\usepackage{amssymb,amsmath}
\def\app#1#2{%
	\mathrel{%
		\setbox0=\hbox{$#1\approx$}%
		\setbox2=\hbox{%
			\rlap{\hbox{$#1\propto$}}%
			\lower1.1\ht0\box0%
			}%
			\raise0.25\ht2\box2%
			}%
			}

\usepackage[hyphens]{url}
\usepackage{revsymb}
\begin{document}
%\sloppy
\title{Jupiter Mass Binaries and Cosmic Ray Viscosity}
\shorttitle{Jupiter Mass Binaries and Cosmic Ray Viscosity}
\shortauthors{Katz}
\author{J. I. Katz\altaffilmark{}}
\affil{Department of Physics and McDonnell Center for the Space Sciences,
Washington University, St. Louis, Mo. 63130}
\email{katz@wuphys.wustl.edu}
%\shortauthors{Katz}
%\date{\today}
%\hypersetup{draft}  %Permits references to break across columns
%\begin{document} %MNRAS
%\psfrag{theta}{$\theta$}  Works only on COMPLETE strings
%\label{firstpage} %MNRAS
%\pagerange{\pageref{firstpage}--\pageref{lastpage}} %MNRAS
%\maketitle %MNRAS
\begin{abstract}
The fraction of planetary mass objects in the Trapezium cluster that are in
wide binaries is much greater than implied by extrapolation to lower masses
of the fraction of stars that are wide binaries.  Wide binaries may be
produced by gravitational collapse of a medium with fluid vorticity.  In a
uniform medium with uniform vorticity the collapse criterion is independent
of the size and mass of the collapsing region, which would imply a wide
binary fraction independent of mass, in contradiction to observation.
Angular momentum, rather than thermal pressure, may be the chief obstacle
to star formation.  The excess of Jupiter Mass Binary Objects in the
Trapezium cluster may be attributed to cosmic ray viscosity that transports
angular momentum to surrounding material.  Viscosity is more effective in
smaller and less massive collapsing regions, preferentially producing
planetary mass wide binaries.  Energetic particles trapped on magnetic field
lines in a corona contribute to viscosity, dissipation and angular momentum
flow in accretion discs.
\end{abstract}
\newpage
\section{Introduction}
\citet{PM23} discovered that an unexpectedly large fraction of Jupiter mass
objects in the Trapezium are in wide (projected separation $\gtrsim
100\,$AU) binaries. At the lowest masses ($M \lesssim 0.002 M_\odot$) this
fraction is $9 \pm 3\%$.  Below about 0.1$M_\odot$ the fraction is an
increasing function of decreasing mass, while above 0.1$M_\odot$ it is an
decreasing function of decreasing mass; at the minimum around 0.1$M_\odot$
the wide binary fraction is $\lesssim 2\%$.  The change in sign of the
dependence on mass around 0.1$M_\odot$ indicates a characteristic mass scale
of wide binary formation, at least in this star- and planet-forming region.

Star and planet formation in the Trapezium cluster is comparatively recent.
These dynamically fragile planetary mass wide binaries with escape
velocities ${\cal O}(100\,\text{m/s}$) have been little affected by
gravitational interaction with other stellar or planetary objects.
Their present-day statistics of orbital parameters provides information
about their processes of formation, information that has been lost from the
statistics of wide field binaries.

Formation from three unbound slow-moving single objects is unlikely because
it is rare that three unbound objects would have such low relative
velocities and be close enough (within approximately the final binary
separation) for two to be captured with expulsion of the third.  That
process would also preferentially produce more massive binaries, in
contradiction to observation, because such three-body interactions
preferentially expel the lowest mass object.  Because of their dynamical
fragility, wide binary planetary mass objects cannot be explained by
expulsion from more tightly bound multiple systems.

The theory of the formation of wide binaries remains controversial and
uncertain, and presents different issues than the formation of close
binaries from the late-stage collapse of a rotating cloud \citep{T02,O22}.
Their formation is attributed to the collapse of a subregion of the
interstellar cloud to a density comparable to the density of the binary's
mass spread over its orbital volume, at which point angular momentum arrests
further collapse on a dynamical time scale.  Unlike field objects
\citep{O22,F23}, the Trapezium wide binaries are a homogeneous population
with recent coeval formation, perhaps from a nearly uniform medium, and
likely have been unaffected by subsequent interactions.  Their binding
energies and masses are orders of magnitude below those discussed by
\cite{JF13} as the result of turbulent fragmentation.  Because of the
complexity of the problem and the heterogeneity of the interstellar medium,
I consider scaling relations, with the objective of explaining the observed
\citep{PM23} increase of the wide binary fraction toward planetary masses.
\section{Angular Momentum}
Angular momentum is a critical obstacle to collapse of a gas cloud
\citep{K87}.  Thermal energy can be radiated away, rapidly at high density,
and magnetic energy turned into thermal energy by resistivity, but angular
momentum cannot be.  The angular momentum of a collapsing cloud almost
entirely appears as the orbital angular momentum of the binary.  Rotation of
the final hydrostatic objects can only take up a tiny fraction (${\cal O}
(10^{-3})$ in the case of binary Jupiters at 100 AU separation) of the
orbital angular momentum.  Here we are concerned with the conditions under
which a subregion of an extended cloud can begin to collapse and the viscous
transport of some of its angular momentum to its surroundings that enables
that collapse.

If a gas cloud collapses to form a single star or planetary mass object,
it is not possible to infer the pre-collapse conditions in the cloud from
the properties of the condensed object; the mass is known, but the
pre-collapse length scale, density and angular momentum are not.  Because
the moment of inertia of a single star or planet is orders of magnitude less
than that of its much larger pre-collapse gas cloud, nearly all of the
pre-collapse angular momentum must have been lost.

The pre-collapse angular momentum of a newly formed binary appears as
orbital angular momentum.  The mean density of the binary's orbital volume
$\langle \rho \rangle \equiv M/(4\pi R^3/3)$, where $M$ is its total mass
and $R$ the binary separation, provides an estimate of the pre-collapse
density.  This follows from the assumption that the cloud's angular momentum
does not stop collapse until it reaches this density, at which point the
cloud fragments into proto-stars or proto-planets whose relative motion
takes up the angular momentum.  If there are more than two fragments, those
of the lowest mass are generally expelled gravitationally, leaving a
surviving binary (\citet{PM23} report 40 binaries but only two triples).

These wide binaries are dynamically fragile, readily disrupted by stellar
encounters, and the observed binary fraction likely underestimates the
fraction at birth.  For a binary of mass $0.002 M_\odot$ ($2M_{Jupiter}$)
and separation 100 AU, like those discovered by \citet{PM23}, the
pre-collapse density of molecular hydrogen $n \sim 10^9\,$cm$^{-3}$.

For a cloud with uniform vorticity ${\vec\omega} \equiv {\vec\nabla}\times
{\vec v}$, where $\vec v$ is the fluid velocity, the velocity difference
$\Delta v_\omega \equiv |{\vec\omega}|R$ over a length $R$ may be compared
to the velocity of gravitational collapse $\Delta v_G \equiv \sqrt{GM/R} =
\sqrt{G 4 \pi \langle \rho \rangle R^2/3}$.  The necessary condition for
collapse of a subregion within a larger homogeneous cloud $\Delta v_\omega
< \Delta v_G$ becomes
\begin{equation}
        \label{condition}
        |{\vec\omega}| < \sqrt{4 \pi G \langle \rho \rangle \over 3}.
\end{equation}
This could be estimated on dimensional grounds, because the express on the
right hand side is the only form containing $G$, $M$, $R$ and $\rho$ with
the dimensions (s$^{-1}$) of vorticity.

Collapse, and formation of a wide binary whose orbit contains the angular
momentum of the collapsing cloud, depends only on the comparison of
$|{\vec\omega}|$ to $\langle \rho \rangle$ in (\ref{condition}).  {These
parameters, averaged over a length scale containing a planetary or stellar
mass of gas, may be approximately uniform through a much larger and more
massive cloud comprising the entire star- and planet-forming region.  The
temperature, set by cosmic ray heating and thermal emission by grains, is
confined to a narrow range, and on these length scales pressure
equilibration is rapid and implies near-uniform density.  This is supported
by the observations of \citet{PM23} of binaries with similar mass densities
over the extended region they observed.}

Conservation of specific angular momentum $\propto VR \propto |{\vec\omega}|
R^2$ implies that as a cloud collapses $|{\vec\omega}| \propto R^{-2}$.  The
characteristic gravitational frequency $\sqrt{4\pi G \langle \rho \rangle/3}
\propto R^{-3/2}$.  At the beginning of collapse the inequality
(\ref{condition}) must be satisfied, but as collapse proceeds $R$ becomes
small enough that (\ref{condition}) is no longer satisfied, and angular
momentum stops collapse.  The likely result is formation of a flattened,
disc-like, structure \citep{O22}.  Because it is self-gravitating, it
separates into two or more denser structures that each collapse into
proto-planets or proto-stars.  If there are more than two such objects
all but two are expelled, and the remaining two, usually the most massive
two, become the components of a wide binary.  \citet{T02} and \citet{O22}
discuss complications of this picture, but these complications do not affect
the control of collapse by angular momentum and the fact that collapse must
stop when the inequality (\ref{condition}) is no longer satisfied.

For the general interstellar medium the rotation of the Galaxy contributes
$|{\vec\omega}| \sim 10^{-15}\,$s$^{-1}$.  For a hydrogen density $n \sim
1\,$cm$^{-3}$ the collapse condition is satisfied by a factor of 5--10.
Comparing the increase $\propto R^{-2}$ on the left hand side of
Eq.~\ref{condition} to that $\propto R^{-3/2}$ on the right hand side shows
that after a decrease in length scale by the square of this factor, and an
increase in density by its sixth power, angular momentum becomes sufficient
to stop collapse.  At least qualitatively, the high power of $R$ may explain
the high inferred density ($n \sim 10^9\,$cm$^{-3}$) of the subregions
forming planetary mass binary objects in the Trapezium cluster.  In fact,
most of the interstellar medium likely has $|{\vec\omega}|$ greater than
that of the general Galactic rotation because the medium is stirred by
supernova shocks, the gravitational fields of spiral arms and massive
clouds, expansion of regions warmed by photoionization and contraction of
those cooled by radiation, and other complex hydrodynamics.
\section{Collapse}
The collapse inequality (\ref{condition}), unlike the Jeans criterion for
collapse of clouds without angular momentum \citep{BT08}, does not depend on
the mass or size of the cloud, and predicts no dependence of the wide binary
fraction on the mass of the binary formed.  It also applies to clouds of
arbitrary size; if the condition is satisfied in a homogeneous cloud, a
large number of wide binaries will be formed, while if not satisfied no
collapse can occur.  Collapse and planetary or star formation occurs when
both criteria are satisfied; the Jeans criterion is satisfied for
Jupiter-mass regions when $n \sim 10^9\,$cm$^{-3}$, consistent with the
pre-collapse density inferred from the present mean density of these
binaries.

In a cloud of uniform density and vorticity the collapse condition
(\ref{condition}) is satisfied either everywhere or nowhere.  But no real
astronomical gas cloud is uniform.   {The vorticity varies in this
complex turbulent flow.  A general clould may comprise regions in which this
condition is satisfied and those in which it is not.}  The density increases
as a result of compressive external pressure, self-gravity, and the
radiative loss of internal energy, and even subregions not initially
satisfying the inequality (\ref{condition}) may come to do so.  The
distribution of masses of these subregions determines the masses of the
objects formed.  If there is no natural ({\it a priori\/}) scale on which
$|{\vec\omega}|$ or $\rho$ vary, then the fraction of wide binaries must be
a power law function of the stellar or planetary mass, the only function
without a characteristic scale.  This is not consistent with the change in
slope of the observed wide binary fraction \citep{PM23} around 0.1$M_\odot$.
That change in slope implies the existence of a characteristic spatial scale
in the cloud.

Collapse begins when (\ref{condition}) is first satisfied; the total mass
of the objects formed is nearly the mass of the gas in the collapsing
region.  Because collapse is limited by angular momentum, the resulting
condensed objects form a binary with separation comparable to the size of
the collapsing subregion when it first satisfies Eq.~\ref{condition}.
Subsequent gravitational interactions with other objects disrupt most of
these wide binaries, and lower mass binaries are particularly susceptible to
disruption because of their weak binding.  Disruption is a random process
and in a very young cloud with active star formation like the Trapezium
cluster many wide binaries are still present, despite the high density of
potentially disrupting stellar or planetary mass objects.
\section{Vorticity}
The observations of \citet{PM23} contradict the prediction, based on the
absence of a characteristic scale of angular-momentum limited collapse
(condition \ref{condition}), that the wide binary fraction should be
independent of $R$ (and hence of $M$) in an initially homogeneous cloud.
This prediction does not apply if $|\langle{\vec\omega}\rangle|$ depends on
the length scale $R$ because, if so, whether a region satisfies the collapse
criterion would depend on its size.  Viscosity diffuses vorticity, with a
characteristic time scale $\propto R^2$.  The evolution of vorticity is
described by
\begin{equation}
        \label{vorticity}
        {d{\vec\omega} \over dt} = ({\vec\omega}\cdot{\vec\nabla}){\vec v}
        -{\vec\omega}({\vec\nabla}\cdot{\vec v}) + {\eta \over \rho}\nabla^2
        {\vec\omega},
\end{equation}
where we have dropped the baroclinic source term
$({\vec\nabla}\rho\times{\vec\nabla}p)/\rho^2$ because the pressure field
$p$ is unknown and likely small compared to $\rho v^2$ (as a result of
radiation of thermal energy), and also ignore body forces because
self-gravity conserves angular momentum.

In a simple (two-dimensional) geometry in which $\vec v$ lies in the $x$-$y$
plane and is independent of $z$, $\vec\omega$ is in the $z$ direction and
the first term on the right hand side is zero.  The most important part of
Eq.~\ref{vorticity} is the second term on the right hand side, which is
large and positive (${\vec\nabla}\cdot{\vec v} < 0$) in a contracting
subregion.  Scaling or dimensional analysis shows that, if viscosity is
unimportant, as the subregion contracts the vorticity grows in proportion to
the density:
\begin{equation}
        \label{vort}
        \begin{split}
                {d{\vec\omega} \over dt} &\approx {\vec\omega} {1 \over \rho}
                {\partial \rho \over \partial t},\\
                |{\vec\omega}| &\propto \rho.
        \end{split}
\end{equation}
As a result, the left hand side of (\ref{condition}) grows faster than the
right hand side until angular momentum prevents further contraction.

If self-gravity is negligible, as is likely for the length scales and masses
considered here ($\sqrt{4 \pi G \rho/3} \sim 2 \times 10^{-11}\,$s$^{-1}$),
viscosity homogenizes the vorticity of an isolated subregion, leaving it in
a state of uniform (rigid-body) rotation.  Contraction may resume if
viscosity diffuses vorticity away from the contracting medium and transports
angular momentum to its less vortical surroundings.  Viscous transport of
vorticity (and angular momentum) from the collapsing subregion competes with
the amplification of vorticity by collapse.

Viscosity is more effective on smaller spatial scales because in
Eq.~\ref{vorticity} it is a diffusive term, reducing the vorticity and
angular momentum of smaller, less massive, subregions more effectively than
those of larger, more massive, regions.  As a result, less massive
subregions more readily contract, increasing the wide binary fraction of
less massive objects.  This scaling may account for the larger wide binary
fraction of the least massive objects discovered by \citet{PM23}.

When self-gravity becomes significant, angular momentum flows from the inner
parts of the cloud to its outer parts, and it tends towards an accretion
disc in which Kepler's laws are inconsistent with uniform vorticity.  In a
Keplerian disc $|{\vec\omega}| \propto r^{-3/2}$.  This occurs later in the
collapse, after most of the initial angular momentum has been transferred to
the orbital angular momentum of the binary proto-planets.
\section{Cosmic Ray Viscosity}
It remains to find a significant source of viscosity.  The atomic viscosity
of gases is negligible on astronomical scales, but interstellar gas contains
cosmic rays that contribute a large effective viscosity because of their
long free paths and relativistic velocities \citep{EJM88}.  The magnetic
fields permeating the gas prevent the free escape of the cosmic rays,
although a quantitative description would require knowledge of the magnetic
field topology---the particles are free to travel along field lines, so
transport in that direction depends on whether they form closed loops and
the dimensions of such loops, and on whether there are magnetic mirrors
along their lengths.

As a rough approximation, the cosmic ray dynamic viscosity
\begin{equation}
        \label{eta}
        \eta \sim n_{CR} (\gamma - 1) m_{CR} v_{CR} \ell_{CR},
\end{equation}
where $n_{CR}$, $m_{CR}$, $\gamma$, $v_{CR} \approx c$ and $\ell_{CR}$ are
respectively the cosmic ray number density, mass, Lorentz factor, speed and
scattering length, integrated over the cosmic ray spectrum.  For the general
interstellar medium $n_{CR} (\gamma-1) m_{CR}c^2 \sim 1\,$eV/cm$^3$
\citep{W98}.

The scattering length $\ell_{CR}$ is strongly anisotropic.  In the
directions perpendicular to the magnetic field it may be estimated (in the
spirit of Bohm diffusion \citep{C84})
\begin{equation}
        \label{Bohm}
        \ell_\perp \sim r_g = {m_{CR}c^2 \over eB},
\end{equation}
where the gyroradius $r_g$ of a cosmic ray proton with 1 GeV kinetic energy
in a representative interstellar $3\,\mu$G magnetic field $r_g \sim 2 \times
10^{12}\,$cm.  If $\ell_{CR}$ is taken as $\ell_\perp$ then $\eta \sim
10^{-10}\,$g/cm-s, which is negligible.

But in the direction parallel to the magnetic field a cosmic ray may travel
a length comparable to the radius of curvature of the field.  We know from
comparing the rotation measures of pulsars to the magnitude of the
interstellar field that the interstellar field is ordered on long scales.
Hence we estimate
\begin{equation}
        \ell_\parallel \sim R,
\end{equation}
where R is the radius of a collapsing subregion; once that subregion has
reached a density a few times higher than that of its surroundings the field
must bend by an angle ${\cal O}(1\,\text{rad})$ between the subregion and
its less dense surroundings.  That bend will not scatter the cosmic rays
(its radius of curvature is $\sim R \gg r_g$), but in being bent the cosmic
rays couple their momentum into the surrounding gas that anchors the
magnetic field and provide an effective viscosity; $R$ is the length on
which the cosmic rays couple momentum.

%The viscosity (most generally described by a fourth-rank tensor,
%like an elastic coefficient, relating the second rank rate of strain tensor
%to the second rank stress tensor) is strongly anisotropic.  A cosmic ray
%with a pitch angle (with respect to the magnetic field) that is neither
%close to zero nor to $\pi/2$ interacts successively with regions displaced
%by ${\cal O}(r_g)$ both parallel and perpendicular to the field, supporting
%the estimate $\ell_{CR} \sim r_g$.  The small fraction of cosmic rays with
%pitch angles close to $\pi/2$ have anisotropic $\ell_{CR}$, small in the
%direction of the field but ${\cal O}(r_g)$ in the other two directions.
%The small fraction of cosmic rays with small pitch angles have $\ell_{CR}$
%in the field direction comparable to the magnetic radius of curvature but,
%provided their transverse momentum is ${\cal O}(m_p c)$, $\ell_{CR} \sim
%r_g$ in the two transverse directions.  Without a detailed knowledge of the
%pitch angle distribution and the field geometry it is not possible to do
%better than taking the Bohm-like estimate Eq.~\ref{Bohm}.

The separations and masses of binary planetary mass objects in the Trapezium
cluster indicate that they formed at a density $\sim 10^9$ times that of
the general interstellar medium.  The column density of the collapsing cloud
$nR \sim 10^{24}\,$cm$^{-2}$ is less than the collisional interaction length
$\sim 3 \times 10^{25}\,$cm$^{-2}$ of cosmic rays with hydrogen, so cosmic
rays may flow through the collapsing cloud, viscously coupling it to its
surroundings.  Because cosmic rays are inextricably bound (by the magnetic
field) to the conducting interstellar gas, their density $n_{CR} \propto n$
as the cloud collapses.  The energy of individual cosmic rays is increased
by compression of the cosmic ray-gas fluid; because cosmic rays are
relativistic $\gamma \propto n^{1/3}$.  Viscosity scales $\eta \propto n
\gamma \ell_\parallel \propto n \gamma R \propto n \propto R^{-3}$, so that
in a proto-JuMBO subregion with $n \sim 10^9\,$cm$^{-3}$, $\eta \sim 0.5
\times 10^5\,$g/cm-s.

The effect of viscosity is quantified by the Reynolds Number
\begin{equation}
        \label{Reynolds}
        \text{Re} \sim {\rho R V \over \eta}
        \sim {\rho R^2 |{\vec\omega}| \over \eta},
\end{equation}
where $V \sim R |{\vec\omega}|$ is a characteristic velocity and $\eta$ is
the (dynamic) viscosity.  In a region forming a wide Jupiter-mass binary
with separation $R \sim 100\,$AU and mass $M \sim 10^{-3}M_\odot$ the
density $\rho \sim 10^{-15}\,$g/cm$^3$.  For a subregion on the threshold of
collapse to a binary planet, estimating $|{\vec\omega}| \sim 2 \times
10^{-11}\,$s$^{-1}$ from (\ref{condition}), leads to a Reynolds number Re
$\sim 1$, implying effective removal of angular momentum from the
subregion to its surroundings.

This Re is small enough to imply that, in clouds with little enough
vorticity to permit collapse to begin, viscosity can transfer enough of
their angular momentum to surrounding matter to permit further collapse to a
wide binary (the binary orbit taking up the remaining angular momentum).
The $R^2$ factor in Eq.~\ref{Reynolds} explains why this process is more
effective for smaller (lower mass) clouds, and leads to formation of a
larger fraction of wide binaries at low masses.  This scaling facilitates
the collapse of smaller clouds more than that of larger clouds, and hence
prefers the formation of less massive wide binaries over more massive wide
binaries, explaining the observations of \citet{PM23}.

In the subsequent collapse of any particular cloud, Re increases $\propto
1/R$ (Eq.~\ref{Reynolds}, noting that $|{\vec\omega}| \propto \rho
\propto R^{-3}$ (Eq.~\ref{vort}) and $\gamma \propto R^{-1}$).  {The
cloud will rotate nearly rigidly, until the increasing importance of gravity
forces its matter into a proto-Keplerian accretion disc.  Cosmic ray
viscosity will continue to transfer angular momentum from the cloud to its
surroundings, facilitating its collapse.}
\section{Discussion}
The maximum $|{\vec\omega}|$ consistent with collapse (\ref{condition})
to form the wide planetary binaries discovered by \citet{PM23} is several
orders of magnitude less than values implied by the scaling relation
$|{\vec\omega}| \propto \rho$ derived from Eq.~\ref{vorticity}.
Star or planet (even wide binary) formation only occurs in subregions
whose complex hydrodynamics has fortuitously led to vorticity much less
than its mean in the extended cloud.  At any moment only a tiny fraction
of the mass of a cloud like that in the Trapezium cluster has sufficiently
low vorticity to form stars or planets.

\citet{PM23} found that the $A_V$ of the primary and secondary objects
typically differ by $\sim 5$ magnitudes.  Combining this with a nominal
interstellar extinction per unit hydrogen column density of
$2\,\text{mag/kpc-cm}^{-3}$ implies a gas density over a 100 AU difference
in distance $n \sim 5 \times 10^6\,$cm$^{-3}$, about 200 times less than the
density required to form the binary.  The intra-binary space must have been
largely cleared of gas since the binary was formed, either by accretion onto
the planetary objects or by expulsion by their changing gravitational
potential.  An alternative explanation is that binaries still immersed in
the very dense pre-collapse gas are difficult to detect, and even more
difficult to recognize as binaries (requiring detection of both objects),
because of their excessive extinction, even in the infrared, so that
observational selection favors detection of binaries in less dense regions.

An implication of the arguments presented here is that nearly all stars (and
free-floating planets) form as wide binaries because of the difficulty and
necessity of removing angular momentum.  Some survive to be observed
as common proper motion binaries, but most are likely disrupted by
gravitational interaction with other objects in their star-forming regions
of birth or subsequently in the interstellar medium.  This predicts an
anticorrelation of wide binary fraction in star-forming regions with these
regions' age and also with their stellar density.

Another implication is that energetic particle viscosity may contribute
to angular momentum transport in other astronomical environments.  It may
occur in the dilute coron\ae\ of accretion discs where relativistic
particles are accelerated, and may contribute significantly to their
viscosity and energy dissipation.  Particles may be trapped on field lines
with feet in the dense disc plasma, whose differential rotation accelerates
the particles in analogy to second-order Fermi acceleration.  This leads to
a mutually amplifying feedback among particle acceleration, radial flow of
mass and angular momentum, and viscous dissipation.  Escape of particles
along open field lines that result from magnetic reconnection powers the
extended radio sources of quasars and microquasars \citep{K24}.
\section*{Compliance with Ethical Standards}
The author has no potential conflicts of interest.
\section*{Data Availability}
This theoretical study generated no new data.
%\bibliography{JuMBOapspsci} %Doesn't work: no \bibstyle command in .aux file
%Use MNRAS format (worked in qso.d/quasarmicroquasarapspsci.tex)

\end{document}